\documentclass[
 reprint,
 amsmath,amssymb,
 prl,
 floatfix,
 superscriptaddress,
]{revtex4-2}
\usepackage[dvipdfmx]{graphicx} 
\usepackage{dcolumn}
\usepackage{bm}
\usepackage{hyperref}
\usepackage{physics}
\usepackage{mathrsfs}
\usepackage{overpic}
\usepackage{bbm}
\usepackage{float}
\usepackage{braket}

\hypersetup{colorlinks = true, 
  urlcolor = blue, linkcolor = blue, citecolor = blue}

\begin{document}

\preprint{APS/123-QED}

\title{Coherent transport in two-dimensional disordered potentials under spatially uniform SU(2) gauge fields}

\author{Masataka Kakoi}
\email{kakoi@presto.phys.sci.osaka-u.ac.jp}
\affiliation{Department of Physics, The University of Osaka, Toyonaka, Osaka 560-0043, Japan}

\author{Christian Miniatura}
\email{christian.miniatura@cnrs.fr}
\affiliation{Universit{\'e} C{\^o}te d'Azur, CNRS, INPHYNI, 17, rue Julien Lauprêtre, 06200 Nice, France}
\affiliation{Centre for Quantum Technologies, National University of Singapore, Singapore}

\author{Keith Slevin}
\email{slevin.keith.sci@osaka-u.ac.jp}
\affiliation{Department of Physics, The University of Osaka, Toyonaka, Osaka 560-0043, Japan}

\date{\today}

\begin{abstract}
    We study interference effects in the dynamics of a spin-$1/2$ particle propagating in two dimensions in a disordered potential and subject to a generalized spin--orbit coupling. 
    With the particle initially in a spin-polarized plane wave state, in the short-time regime, before the spin and momentum distributions reach their steady states, we observe a transient backscattering peak offset from the exact backscattering direction, coexisting with a coherent backscattering dip. 
    We present an intuitive explanation of this momentum offset using a non-Abelian gauge transformation. 
    We also describe the full time evolution of the transient peak, from its buildup to its decay with a precise prediction of the dephasing time within a perturbative framework for multiple scattering.
    Our results can be applied to general spatially uniform SU(2) gauge fields, including the synthetic gauge field in ultracold atoms. 
\end{abstract}

\maketitle

Waves traveling along reciprocal paths in disordered media may interfere constructively, leading to an interference effect known as coherent backscattering (CBS) that has been observed for various waves~\cite{Kuga_1984,VanAlbada_1985,*Wolf_1985,Labeyrie_1999,Bayer_1993,Larose_2004,Jendrzejewski_2012a,Labeyrie_2012,Hainaut_2017}. 
CBS is sensitive to perturbations that break reciprocity~\cite{Sigwarth_2022_review}, such as external magnetic fields, which introduce an additional phase on the wave function. 
In solid state physics, the related phenomena of weak localization are observed in 
magnetoresistance measurements~\cite{Bergmann_1984_review,PA-Lee_1985_review} that are widely used to probe quantum interference effects in solids.
Echo phenomena in response to dephasing pulses have also been studied~\cite{Micklitz_2015,Muller_2015,ZL-Li_2024}.

Over the past decade, ultracold atoms have developed into an important platform for the study of disordered and quasiperiodic systems, as they allow for experimental control of interactions and spatial dimensionality.
The small energy scales of cold atoms allow for direct, time-resolved observation, of density distributions in both real and momentum space.
This has prompted studies of the effects of CBS, focusing on out-of-equilibrium time evolution~\cite{Cherroret_2012,Scoquart_2020,Kakoi_2024,Arabahmadi_2024,Cherroret_2021_review,Jendrzejewski_2012a,Labeyrie_2012,Hainaut_2018}.

With spin--orbit coupling (SOC), which breaks spin-rotation symmetry while preserving time-reversal symmetry, the interference between different spin channels becomes destructive, and CBS is reduced. 
In solids, this leads to weak anti-localization~\cite{Hikami_1980,Bergmann_1984_review}. 
Weak anti-localization can also emerge from general couplings of internal degrees of freedom, such as in graphene~\cite{T-Ando_1998a,*T-Ando_1998b,McCann_2006,Y-Ando_2013_review}. 
SOC can be interpreted as a non-Abelian gauge field and can be emulated across various platforms~\cite{Bliokh_2015_review,Y-Yang_2024_review,Ma_2016,Rechcinska_2019,Y-Chen_2019,Polimeno_2021,Y-Li_2022,J-Wu2022}, including cold atoms~\cite{Galitski_2013_review,Ruseckas_2005,XJ-Liu_2009,Campbell_2011,YJ-Lin_2011,P-Wang_2012,*Cheuk_2012,Huang_2016,Z-Wu_2016,Leroux_2018,Hasan_2022,ZY-Wang_2021,Q-Liang_2024,Madasu_2025}.
With the advent of experimentally tunable non-Abelian gauge fields, the complex wave interference and transport phenomena arising from the interplay of disorder and general gauge fields have attracted attention.
In this Letter, we consider a model that incorporates arbitrary spatially uniform SU(2) gauge fields and a two-dimensional (2D) random potential, and investigate the momentum- and time-resolved transport dynamics. We report the emergence of a transient peak offset from the exact backscattering direction, which coexists with a robust CBS dip. 

{\it Hamiltonian for uniform SU(2) gauge fields---}We consider a spin--orbit coupled particle~\footnote{We use the term ``spin--orbit coupling'' for convenience, though general couplings between momentum and internal degree of freedom, such as sublattice and polarization, are included.} evolving in $d$ spatial dimensions, described by the clean Hamiltonian 
\begin{align}\label{eq:original_Ham}
    \hat{H}_0 = \frac{(\hat{\bm{p}}+\!\hat{\bm{A}})^2}{2m},
\end{align}
where $\hat{\bm{A}}$ is any uniform non-Abelian SU(2) vector potential. As such, we have 
\begin{equation}\label{eq:def_A}
    \hat{\bm{A}} = M\hat{\bm{\sigma}},
\end{equation}
where $M$ is a $d\times3$ real matrix and $\hat{\bm{\sigma}} = (\hat{\sigma}_1,\hat{\sigma}_2,\hat{\sigma}_3)$ denotes the usual set of Pauli spin operators, $\ket{\uparrow,\downarrow}$ being the spin eigenstates of $\hat{\sigma}_{3}$. 
Since $\bm{A}^2/(2m) = \Tr(M M^\top )/(2m) \, \mathbbm{1}$, where $\top$ denotes matrix transposition, the natural energy scale of the system is set by $E_{\kappa} = \hbar^2 \kappa^2/(2m) := \Tr(M M^\top )/(2m)$. In turn, the natural scales for momentum, length, and time are  $\hbar\kappa$, $L_\kappa=1/\kappa$, and $\tau_\kappa=\hbar/E_\kappa$, respectively. Shifting the origin of energies, $\hat{H}_0$ can be recast into the familiar form $\hat{\bm{p}}^2/(2m)  + \hat{\bm{p}}\cdot\!\hat{\bm{A}}/m$.

For $d=2$~\footnote{More generally, for a spin 1/2 particle moving in $d$ spatial dimensions, the number of singular values of $M$ would be $\min(d,3)$. Thus, for $d\ge 3$, and in an appropriate frame, the corresponding gauge field would only have at most 3 nonzero spatial components, one proportional to $\hat{\sigma}_1$, the second to $\hat{\sigma}_2$ and the last to $\hat{\sigma}_3$, the other components being zero~\cite{SM}.}, two parameters are sufficient to characterize $\hat{\bm{A}}$.
That is, regardless of the choice of $M$, there exist suitable rotations of the spatial and spin axes that bring the SU(2) gauge field to the form given as
\begin{align}\label{eq:def_A_equiv}
    \hat{A}_x = \kappa_x \hat{\sigma}_1 = \kappa \cos\eta\,\hat{\sigma}_1,\quad
    \hat{A}_y = \kappa_y \hat{\sigma}_2 = \kappa \sin\eta\,\hat{\sigma}_2,
\end{align}
where $\kappa\geq0$ and $\eta\in[0,\pi/4]$~\cite{SM}.\nocite{Winkler_2004,Fuchs_2010}
In the context of cold atom experiments, the momentum $\hbar{\bm \kappa} = \hbar \,(\kappa_x,\kappa_y)$ is analogous to the recoil momentum, while $E_\kappa$ is akin to the recoil energy~\cite{Hasan_2022}.
The vector potential~(\ref{eq:def_A_equiv}) is well known in the solid-state physics context. For example, pure Rashba~\cite{Bychkov_1984} and pure Dresselhaus~\cite{Dresselhaus_1955} couplings correspond to $\eta=\pi/4$, while $\eta=0$ corresponds to equal strengths of Rashba and Dresselhaus couplings, see Supplemental Material (SM)~\cite{SM}. Note, however, that in this latter case, the gauge field can be gauged away (it has only one component proportional to $\hat{\sigma}_1$). As another example, the same vector potential appears in quantum dots~\cite{Aleiner_2001}.
This equivalence with Eq.~(\ref{eq:def_A}) is important for comparing experiments from the cold atoms, optics and condensed matter communities.
Diagonalizing the Hamiltonian in spin space yields two energy branches (labeled by $\pm$)  
\begin{equation}
    E_{\pm}(\bm{k}) = \frac{\hbar^2k^2}{2m} \pm \frac{\hbar^2\kappa k}{m} \sqrt{\frac{1 + \cos2\eta\cos2\theta}{2}} + E_{\kappa},
    \label{eq:EigenE}
\end{equation}
where $\bm{k}=\bm{p}/\hbar$ is the wave vector associated to momentum $\bm{p}$ and where we used the polar coordinates $k_x = k\cos\theta$, $k_y = k\sin\theta$. 
Throughout this Letter, unless specified otherwise, we fix $\eta=\pi/24$.

\begin{figure}[t]
\begin{center}
    \includegraphics[width=0.95\linewidth]{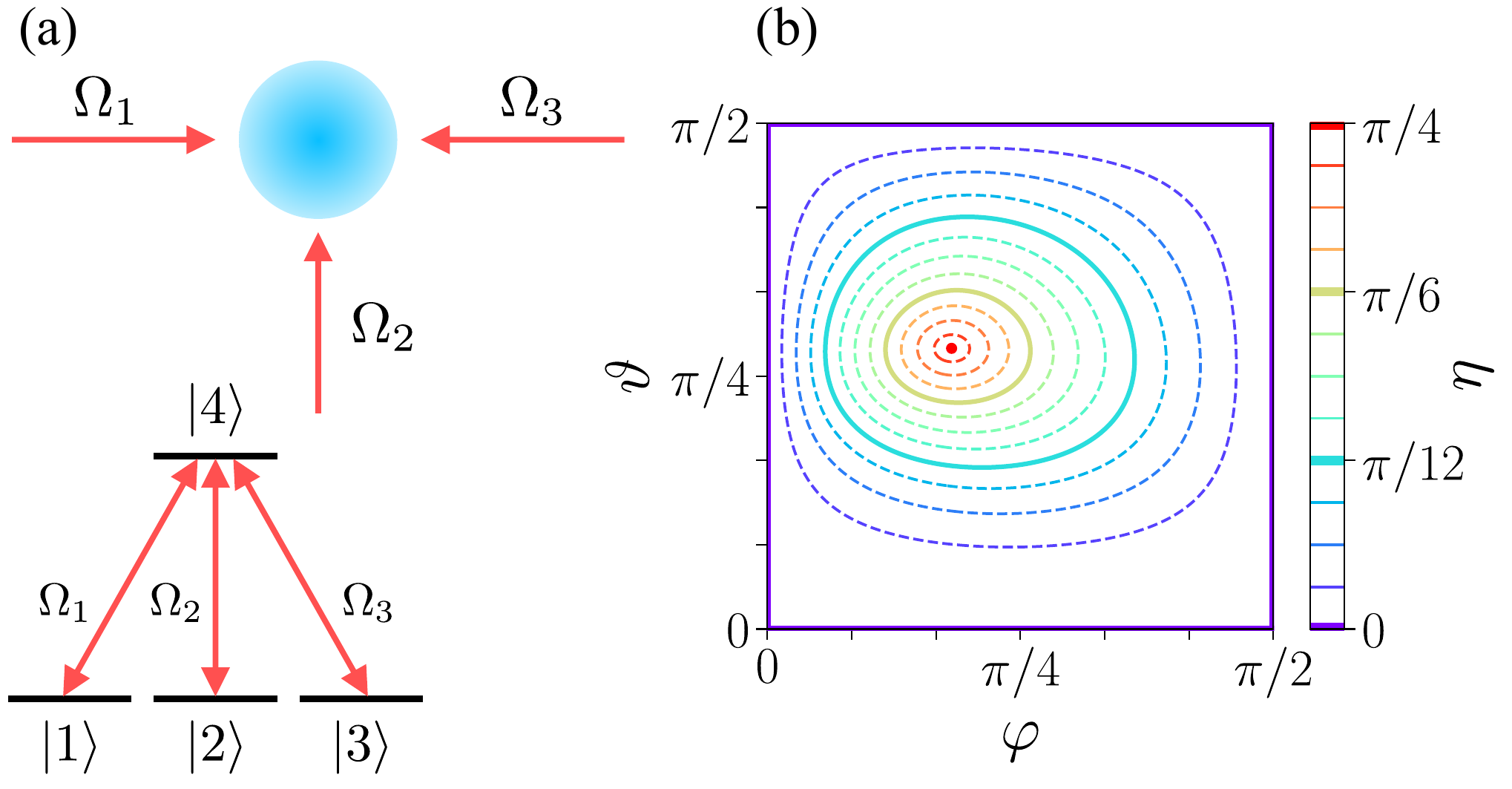}
    \caption{
    (a)~Real-space configuration of the tripod laser scheme used to address a cold atomic cloud (in blue) in Ref.~\cite{Hasan_2022}. The laser beam with Rabi frequency $\Omega_i$ addresses the atomic transition $\ket{i}\leftrightarrow\ket{4}$ ($i=1,2,3$). 
    (b)~$\eta$ values obtained by varying the spherical angles $\vartheta$ and $\varphi$ (see main text).
    The particular choice $\tilde{\bm{\Omega}}= (\cos\varphi \, \sin\vartheta, \sin\varphi \, \sin\vartheta, \cos\vartheta) = (\sqrt{a},a,\sqrt{a})$, where $a=\sqrt{2}-1$ ($\varphi\approx0.57\,\text{rad}$, $\vartheta\approx0.87\,\text{rad}$), leads to $\eta=\pi/4$. Note that $\tilde{\bm{\Omega}}$ is indeed a unit vector since $a^2+2a=1$.
    }
    \label{fig:tripod_eta}
\end{center}
\end{figure}

The entire range of $\eta$ is, in principle, accessible with a cold atomic synthetic SOC created by shining 3 laser beams arranged in a tripod configuration onto the atoms~\cite{Ruseckas_2005,Leroux_2018,Hasan_2022}. Such a system admits two dark states $\ket{D_i}$ ($i=1,2$) and its (adiabatic) dynamics within the dark state manifold is described by Eq.~(\ref{eq:original_Ham}) where the effective SU(2) gauge field is given by $\bm{A}_{ij} =-i\hbar\braket{D_i|\nabla D_j}$~\cite{Ruseckas_2005}. 
For the laser configuration realized in Ref.~\cite{Hasan_2022} [see Fig.~\ref{fig:tripod_eta}(a)], the space-dependent Rabi frequencies read as $\Omega_1 \, e^{ik_{\rm L}x}$, $\Omega_2 \, e^{ik_{\rm L}y}$, and $\Omega_3 \, e^{-ik_{\rm L}x}$ where $k_{\rm L}$ is the laser wave number. 
We next use spherical coordinates ($\Omega_0,\vartheta,\varphi$) to parametrize the Rabi vector $\bm{\Omega} =(\Omega_1,\Omega_2,\Omega_3) = \Omega_0 \, \tilde{\bm{\Omega}}$, where $\Omega_0=\sqrt{\Omega^2_1+\Omega^2_2+\Omega^2_3}$ and $\tilde{\bm{\Omega}} = (\cos\varphi\sin\vartheta, \sin\varphi\sin\vartheta, \cos\vartheta)$. 
Since the dark states depend on the laser Rabi frequencies, so does the gauge field and the corresponding value of $\eta$ can be tuned by varying $\varphi$ and $\vartheta$. As shown in Fig.~\ref{fig:tripod_eta}(b), $\eta$ can be varied from $0$ to $\pi/4$, see SM for details~\cite{SM}.

\begin{figure*}[t]
\begin{center}
    \includegraphics[width=0.9\linewidth]{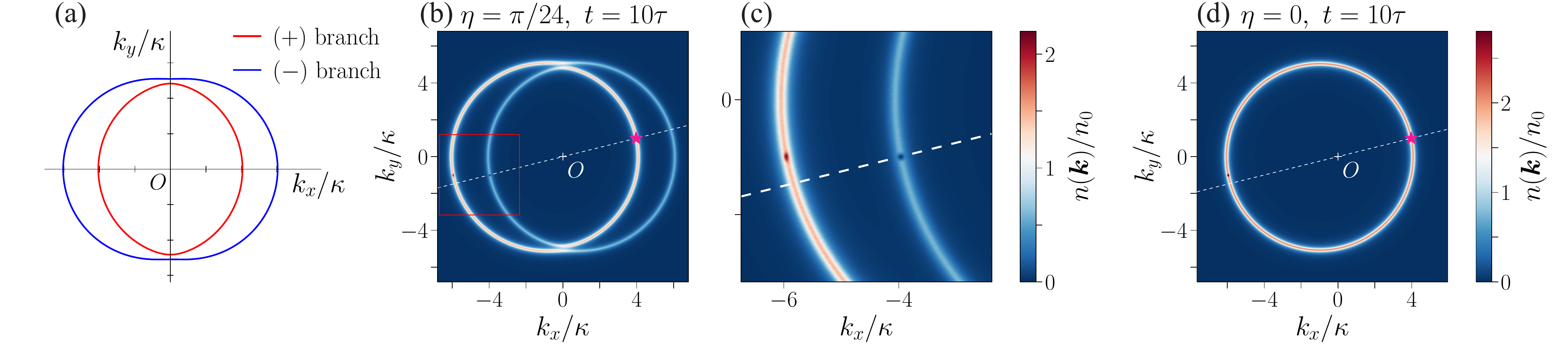}
    \caption{ 
    (a)~The $\pm$ branches $k_\pm(\theta)$ [Eq.~(\ref{eq:FermiB})] of the Fermi surface  for  $\eta=\pi/24$ and $ E_0 \simeq 25.6\, E_\kappa$. Here $\theta$ is the polar angle in the ($k_x,k_y$) plane.
    (b)~The disorder-averaged momentum distribution $n(\bm{k},t)$ at $t=10\tau$ normalized by $n_0=1/(\pi^2\rho^2\gamma_0)$. The initial wave vector $\bm{k}_0$ on the ($+$) branch is indicated by the red star.
    The dashed line connects $\bm{k}_0$ and $-\bm{k}_0$.
    (c)~An enlargement of $n(\bm{k},t)$ in the region enclosed by the red square in (b). A dip is found at $-\bm{k}_0$ on the ($+$) branch and a peak is observed on the ($-$) branch.
    (d)~A comparison with $n(\bm{k},t)$ for $\eta=0$. In this case,  only the peak is observed. Other parameters are the same as in (b). } 
    \label{fig:moment_distrib}
\end{center}
\end{figure*}

We consider now a spin-independent and spatially $\delta$-correlated disordered potential $V(\hat{\bm{r}})$ satisfying
\begin{equation}\label{eq:delta-correlation}
    \overline{V(\bm{r})V(\bm{r}')} = \gamma_0 \, \delta(x-x')\delta(y-y'),
\end{equation}
where $\overline{(\cdot)}$ denotes the average over disorder realizations.
Spin and momentum are coupled, so even a spin-independent disorder affects the spin, and not just the momentum, dynamics.
For $\delta$-correlated disorder, without SOC, scattering is isotropic, whereas with SOC, the scattering probability gains angular dependence from the overlap integral between spin states.

{\it Numerical simulations---}Using the split-step method \cite{Weideman_1986}, we compute the disorder-averaged momentum distribution $n(\bm{k},t):=\overline{|\bra{\bm{k}}e^{-i\hat{H}t}\ket{\bm{k}_0}|^2}$ at time $t$ for an initial plane wave state $\ket{\bm{k}_0}$. $n(\bm{k},t)$ is normalized to satisfy $\int\!\frac{d^2\bm{k}}{(2\pi)^2} n(\bm{k},t) = 1$ and we average over $4000$ disorder realizations.
Within the Born approximation for the self-energy, the scattering mean free time is given by $\tau=\hbar/(\pi\rho\gamma_0)$, where $\rho$ is the total density of states per unit area (DOS) at the initial energy considered~\cite{SM}. 
In the following, we fix $\gamma_0 = 2E_{\kappa}^2L_{\kappa}^2$. 

In Fig.~\ref{fig:moment_distrib}(b), we show $n(\bm{k},t)$ at $t=10\tau$ when the initial state has wave vector $\bm{k}_0=4\kappa_x (1,1/4)$ and is on the $(+)$ branch of the energy dispersion.
The initial energy is $E_0=E_{+}(\bm{k}_0) = \varepsilon_0 E_{\kappa}$ with $\varepsilon_0 \simeq 25.6$. 
For these parameters, the DOS is very close to its clean value in the absence of SOC, i.e., $m/(\pi\hbar^2)$. Thus, $\rho \simeq 1/(2\pi) E_{\kappa}^{-1}L_{\kappa}^{-2}$, $\tau \simeq \tau_{\kappa}$, the scattering mean free path is $\ell \simeq 10\,L_{\kappa}$, and $E_0\tau/\hbar \simeq \varepsilon_0 \gg 1$.
In this weak-disorder regime, the dynamics unfolds on-shell, meaning scattering couples only momentum states with energies  $\approx E_0$. 
These states belong to the two branches $k_\pm (\theta)$ of the Fermi surface, given by
\begin{equation}
\label{eq:FermiB}
    \frac{k_\pm (\theta)}{\kappa} =\sqrt{\frac{2\varepsilon_0 \!-\!1+\cos2\eta \cos2\theta}{2}} \mp \sqrt{\frac{1+\cos2\eta \cos2\theta}{2}},
\end{equation}
and shown in Fig.~\ref{fig:moment_distrib}(a), that are obtained by slicing the energy dispersion $E_{\pm}(\bm{k})$ at $E_0$.
Starting from point $\bm{k}_0$ on the ($+$) branch, at short times $t \sim \tau$, the dynamics gradually fills both branches [Fig.~\ref{fig:moment_distrib}(b)].
At long times $t\gg \tau$, the system enters the ergodic regime where both branches are uniformly filled, except for robust interference effects [Fig.~\ref{fig:moment_distrib}(c)].
We compare with the dynamics when $\eta=0$ in Fig.~\ref{fig:moment_distrib}(d).
In this case, scattering fills only the ($+$) branch with $k_x \ge 0$ and the ($-$) branch with $k_x \le 0$~\cite{SM}.
This reflects an SU(2) spin rotation symmetry which gives rise to a persistent spin helix~\cite{Bernevig_2006,Schliemann_2003,*Schliemann_2017_review,Koralek_2009}.
Weak anti-localization is then absent, leaving room for the observation of weak localization effects~\cite{Pikus_1995,Weigele_2020}.

Looking at the dynamics along the ($+$) branch in Fig.~\ref{fig:moment_distrib}(c), we find a robust CBS dip appearing at $-\bm{k}_0$. 
For Hamiltonians that commute with a time-reversal operator $\hat{\mathcal{T}}$ with $\hat{\mathcal{T}}^2=-1$, the transition amplitudes between time-reversed states are exactly zero at all times, irrespective of the disorder configuration~\cite{Kakoi_2024,Arabahmadi_2024}. 
States at $\pm\bm{k}_0$ on the same branch of the Fermi surface are precisely related by time-reversal symmetry. 
This destructive interference effect reflects a $\pi$ Berry phase acquired by the wave function when circling the Fermi surface, see SM~\cite{SM}. 
A similar CBS dip has been predicted in graphene-like systems subject to correlated random potentials ~\cite{T-Ando_1998a,*T-Ando_1998b,KL-Lee_2014a}. 

Surprisingly, we observe a peak on the ($-$) branch at a momentum offset from the backscattering direction. This is reminiscent of CBS experiments with light when reciprocity is broken~\cite{Lenke_2000a,*Lenke_2000b,Bromberg_2016}.
This peak is transient and decays on a timescale comparable to the equilibration time of the 
diffusive backgrounds across both branches.
Below, we present an intuitive picture that explains the momentum offset. We also reproduce the peak with a diagrammatic calculation and estimate the timescale of the decay. 

{\it Non-Abelian gauge transformation---}
The momentum offset at which the transient peak appears can be qualitatively understood using a non-Abelian gauge transformation that involves a change to a new frame of reference in which the $\hat{p}_x\hat{A}_x/m$ term in Eq.~(\ref{eq:original_Ham}) is removed.
As in Refs.~\cite{Aleiner_2001,Bernevig_2006} in the condensed matter context,
we introduce a spin-dependent unitary operator 
\begin{align}
    \hat{U} &= e^{i\kappa_x \hat{x}\hat{\sigma}_1} = \cos\!\big(\kappa_x\hat{x}\big) +i \sin\!\big(\kappa_x\hat{x}\big)\hat{\sigma}_1.
\end{align}
Contrary to Ref.~\cite{Aleiner_2001} where $L_{\kappa} \gg L$ is assumed ($L$ is the system size), a weak SOC expansion cannot be used here since $L_\kappa \ll \ell \,(\ll L)$ in our case.
Next, we perform the transformation 
\begin{align}\label{eq:transformed_Ham}
    \hat{U}\hat{H}_0\hat{U}^\dag 
    &= \frac{(\hat{\bm{p}}+\!\hat{\bm{A}}')^2}{2m} = \frac{\hat{\bm{p}}^2}{2m} + \hat{H}_{\!\bm{A}} +\frac{(\hat{\bm{A}}')^2}{2m},
\end{align}
where
\begin{equation}\label{eq:A_xy_primed}
    \hat{A}_x' = 0, \quad \hat{A}_y' = \hbar\kappa_y[\cos(2\kappa_x \hat{x}) \, \hat{\sigma}_2 - \sin(2\kappa_x \hat{x}) \, \hat{\sigma}_3].
\end{equation}
Since $(\hat{A}_y')^2=\hbar^2\kappa_y^2$, the final term of Eq.~(\ref{eq:transformed_Ham}) is constant and not essential to the discussion, and
\begin{align}
    \hat{H}_{\!\bm{A}} 
    = \int\!\!\frac{d^2\bm{k}}{(2\pi)^2} \frac{\hbar^2 \kappa_y k_y}{m} 
   \, \ket{\bm{k}\!+\!\bm{\kappa}_x, \rightarrow}\bra{\bm{k}\!-\!\bm{\kappa}_x, \leftarrow}
    + {\rm H.c.}
\end{align}
where $\ket{\bm{k}}$ denotes momentum states, $\bm{\kappa}_x = \kappa_x\,\va{e}_{x}$, and $\ket{\rightleftarrows}$ denote the spin states
\begin{equation}\label{eq:updn2UpDn}
\ket{\rightarrow} = \frac{\ket{\uparrow}+\ket{\downarrow}}{\sqrt{2}}, \qquad \ket{\leftarrow} = \frac{\ket{\uparrow}-\ket{\downarrow}}{\sqrt{2}}.
\end{equation}
The random potential $V(\hat{\bm{r}})$ is not changed by the transformation since it is spin-independent.

\begin{figure}[t]
\begin{center}
    \includegraphics[width=\columnwidth]{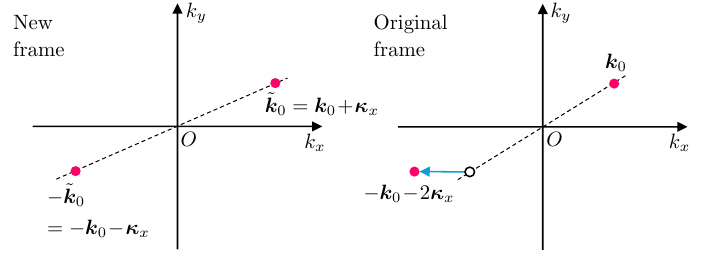}
    \caption{ 
    New frame: For $\eta=0$, the transformed Hamiltonian Eq.~(\ref{eq:transformed_Ham}) is diagonal in spin. The initial state is $\ket{\tilde{\bm{k}}_0, \rightarrow}$. Disorder induces a CBS peak at $\ket{-\tilde{\bm{k}}_0, \rightarrow}$. Original frame: The CBS peak is offset from $-\bm{k}_0$ by $-2{\bm \kappa}_x$, see Fig.~\ref{fig:moment_distrib}(d).
    } 
    \label{fig:shift_schematic}
\end{center}
\end{figure}

We consider first $\eta=0$. Then, $\hat{H}_{\!\bm{A}}=0$ and the transformed Hamiltonian, which reduces to $\hat{\bm{p}}^2/(2m)+V(\hat{\bm{r}})$, is diagonal in spin. The spin states $\ket{\rightleftarrows}$ are dynamically decoupled and the time reversal properties of the transformed Hamiltonian are those of spinless particles with $\hat{\mathcal{T}}^2=1$. Disorder now induces a robust CBS peak rather than a dip. In the new frame, the initial state is $\ket{\tilde{\bm{k}}_0, \rightarrow}$, where $\tilde{\bm{k}}_0 = \bm{k}_0 + {\bm \kappa}_x$, and the peak appears at $-\tilde{\bm{k}}_0$. In the original frame, the peak is at $-{\bm k}_0 - 2{\bm \kappa}_x$ (Fig.~\ref{fig:shift_schematic}). 

When $\eta$ is nonzero but small, the short-time dynamics is still predominantly governed by $\hat{\bm{p}}^2/2m + V(\hat{\bm{r}})$.
The position of the CBS peak remains approximately at $-\bm{k}_0- 2\bm{\kappa}_x$. For the specific initial wave vector chosen, an elementary geometrical calculation gives the shift angle $\delta\theta\approx 0.08$ radians from the backscattering direction, a value confirmed by our numerical observations in Fig.~\ref{fig:moment_distrib}(b)-(d). 
However, since $\eta\neq 0$, $\hat{H}_{\!\bm{A}}$ couples the spin states and the relevant time reversal operator has $\hat{\mathcal{T}}^2=-1$. 
This leads to dephasing that disrupts the constructive interference that causes the CBS peak, and the peak decays.
For small $\eta$,  second-order perturbation theory suggests that the dephasing rate scales with the square of the coupling strength, i.e., like $\kappa^2_y \propto \sin^2\!\eta \approx \eta^2$~\cite{SM}.

{\it Diagrammatic calculation for the time evolution---}To quantitatively predict the time evolution of the transient peak in the weak-disorder regime, including its lifetime, we present an analytical calculation based on Refs.~\cite{Kuhn_2005,*Kuhn_2007,Cherroret_2012,Scoquart_2020,Kakoi_2024}, using the maximally-crossed diagram series (Cooperon).
At $\eta=0$,  the disordered Hamiltonian $\hat{\bm{p}}^2/(2m)+V(\hat{\bm{r}})$ is block-diagonal in the spin space spanned by $\ket{\rightleftarrows}$. 
At small $\eta$, we expect $\hat{H}_{\!\bm{A}}$, which couples the two spin states with a strength proportional to $\sin^2\!\eta \approx \eta^2$, to act as a perturbation. 
In this case, the off-diagonal components of the Cooperon in the $\ket{\rightleftarrows}$ basis are negligible, at least at short times.
In this approximation, the Cooperon in the $\ket{\rightarrow}$ subspace is calculated from the Bethe--Salpeter equation:
\begin{equation}
    \Gamma^{\mathrm{C}}_{\rightarrow}(\bm{q},\omega,E) = \frac{\gamma_0}{1 - \gamma_0\Pi_{\rightarrow}^C(\bm{q},\omega,E)},
\end{equation}
where the propagator $\Pi^{\mathrm{C}}_{\rightarrow}$ is given by
\begin{align}
\label{eq:PropPiC}
    \Pi^{\mathrm{C}}_{\rightarrow}(\bm{q},\omega,E) =\int\!\!\frac{d^2\bm{k}}{(2\pi)^2} 
    \overline{G}_{\rightarrow}\!\left(\bm{k},E\right)
    \overline{G}_{\rightarrow}^*\!\left(\bm{q}\!-\!\bm{k},E\!-\!\hbar\omega\right).
\end{align}
Here, $\overline{G}_{\rightarrow}(\bm{k},E) = \bra{\rightarrow}\overline{\hat{G}}(\bm{k},E)\ket{\rightarrow}$ is the disorder-averaged Green's function
\begin{equation}\label{eq:GF_rightleftarrows}
    \overline{G}_{\rightarrow}(\bm{k},E) =\frac{\cos^2\big(\phi_\eta(\theta)/2\big)}{E-E_+(\bm{k})+ i\frac{\hbar}{2\tau}} + \frac{\sin^2\big(\phi_\eta(\theta)/2\big)}{E-E_-(\bm{k})+ i\frac{\hbar}{2\tau}},
\end{equation}
where $\phi_\eta(\theta) = \arg(\cos\eta\cos\theta + i\sin\eta\sin\theta)$ also appears in the phase of the eigenstates of $\hat{H}_0$.

The contribution from the Cooperon at wave vector $\bm{k}$, given an initial wave vector $\bm{k}_0$, is characterized by $\Gamma^{\mathrm{C}}_{\rightarrow}(\bm{k}+\bm{k}_0)$.
The small $\omega$ expansion of $\Pi^{\mathrm{C}}_{\rightarrow}$ around $\bm{q}=-2\bm{\kappa}_x$ at $E=E_0$ gives 
\begin{equation}\label{eq:expand_propagator}
    \gamma_0\Pi^{\mathrm{C}}_{\rightarrow}(\bm{q},\omega,E_0) = 1+i\tau\omega - D\tau (\bm{q}\!+\!2\bm{\kappa}_x)^2 - \frac{\tau}{\tau_\gamma},
\end{equation}
where $D=v_g\ell/2$ is the diffusion constant ($v_g$ is the initial group velocity) and $D\tau = \ell^2/2$. Here, 
\begin{equation}\label{eq:lifetime}
    \tau_\gamma = \frac{\tau}{1 - \gamma_0\Pi^{\mathrm{C}}_{\rightarrow}(-2\bm{\kappa}_x,0,E_0)}
\end{equation}
defines the dephasing time $\tau_\gamma$. 
When $\eta=0$ we find $\tau_\gamma=\infty$ as expected because, $\hat{H}_{\bm{A}}=0$, Eq.~(\ref{eq:GF_rightleftarrows}) becomes
\begin{equation}
    \overline{G}_{\rightarrow}(\bm{k},E) = \Big[E-\frac{\hbar^2(\bm{k}\!+\!\bm{\kappa}_x)^2}{2m}+i\frac{\hbar}{2\tau}\Big]^{-1},
\end{equation}
entailing $\gamma_0 \Pi^{\mathrm{C}}_{\rightarrow}(-2\kappa \va{e}_x,0,E_0)=1$ via Eq.~(\ref{eq:PropPiC}).
From the $\omega$- and $\bm{q}$-dependence of $\Pi^{\mathrm{C}}_{\rightarrow}$, through $\Gamma^{\mathrm{C}}_{\rightarrow}$, the time dependence of the contribution $n^{\mathrm{C}}(\bm{k},t)$ of the Cooperon to the momentum distribution can be computed using the residue theorem, see Refs.~\cite{Scoquart_2020,Kakoi_2024}.

\begin{figure}[t]
\begin{center}
    \includegraphics[width=0.95\columnwidth]{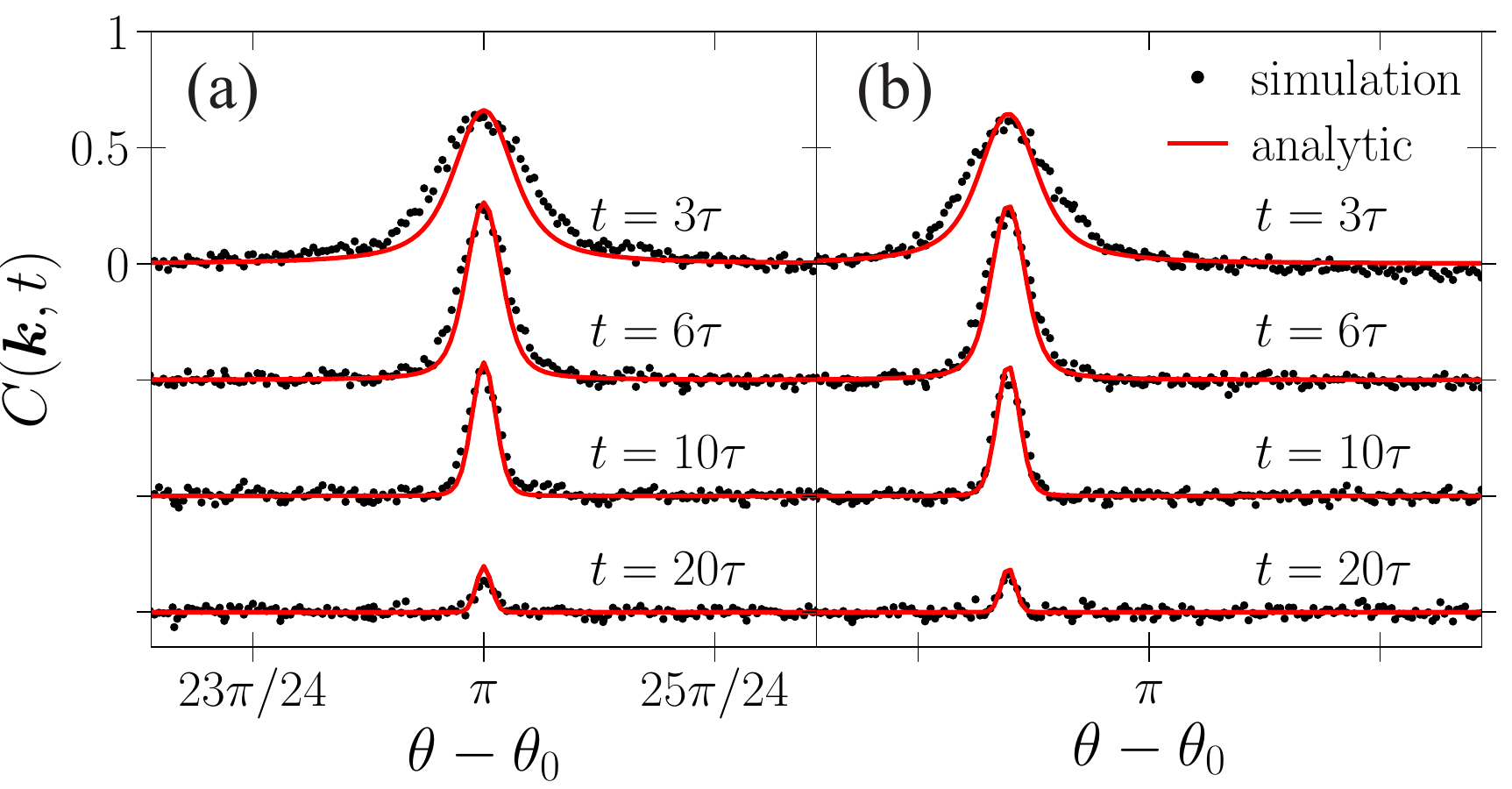}
    \caption{
    For $\eta=\pi/24$, (black) the peak to background ratio $C(\bm{k},t)$ 
    obtained in  simulations, and (red) the ratio $n^{\mathrm{C}}(\bm{k},t) / n_{\mathrm{B}}(\bm{k},t)$, at the times indicated. The initial wave vector is $\bm{k}_0 = 4\bm{\kappa}_x$ in (a) and $\bm{k}_0=\kappa_x (4\va{e}_{x}+\va{e}_{y}/2)$ in (b). $\theta$ and $\theta_0$ denote the polar angles of $\bm{k}$ and $\bm{k}_0$, respectively. Plots for different times have been offset for clarity.
    } 
    \label{fig:simulation_analytic_calc}
\end{center}
\end{figure}

For detailed quantitative analysis we focus on the peak-to-background ratio 
\begin{equation}\label{eq:peak-to-background}
    C(\bm{k},t)  =  n_{\mathrm{I}}(\bm{k},t) / n_{\mathrm{B}}(\bm{k},t).
\end{equation}
Here, $n_{\mathrm{B}}(\bm{k},t)$ is the smooth diffusive background and
\begin{equation}
    n_{\mathrm{I}}(\bm{k},t) = n(\bm{k},t) - n_{\mathrm{B}}(\bm{k},t)
\end{equation}
is the contribution due to interference effects. 
In Fig.~\ref{fig:simulation_analytic_calc}, we show $C(\bm{k},t)$ obtained from the simulation,
where for $n_{\mathrm{B}}(\bm{k},t)$ we have taken the numerical value of the smooth flat background around the CBS peak. 
The data are well reproduced, without any adjustable parameters, by the ratio 
\begin{equation}
    n^{\mathrm{C}}(\bm{k},t) / n_{\mathrm{B}}(\bm{k},t)
\end{equation}
with the Cooperon obtained from Eq.~(\ref{eq:expand_propagator}).
The full analytic calculation of the time evolution of the momentum distribution $n(\bm{k},t)$, including the CBS dip and the diffusive background, will be presented elsewhere~\cite{Kakoi_2026}.

\begin{figure}[t]
\begin{center}
    \includegraphics[width=\columnwidth]{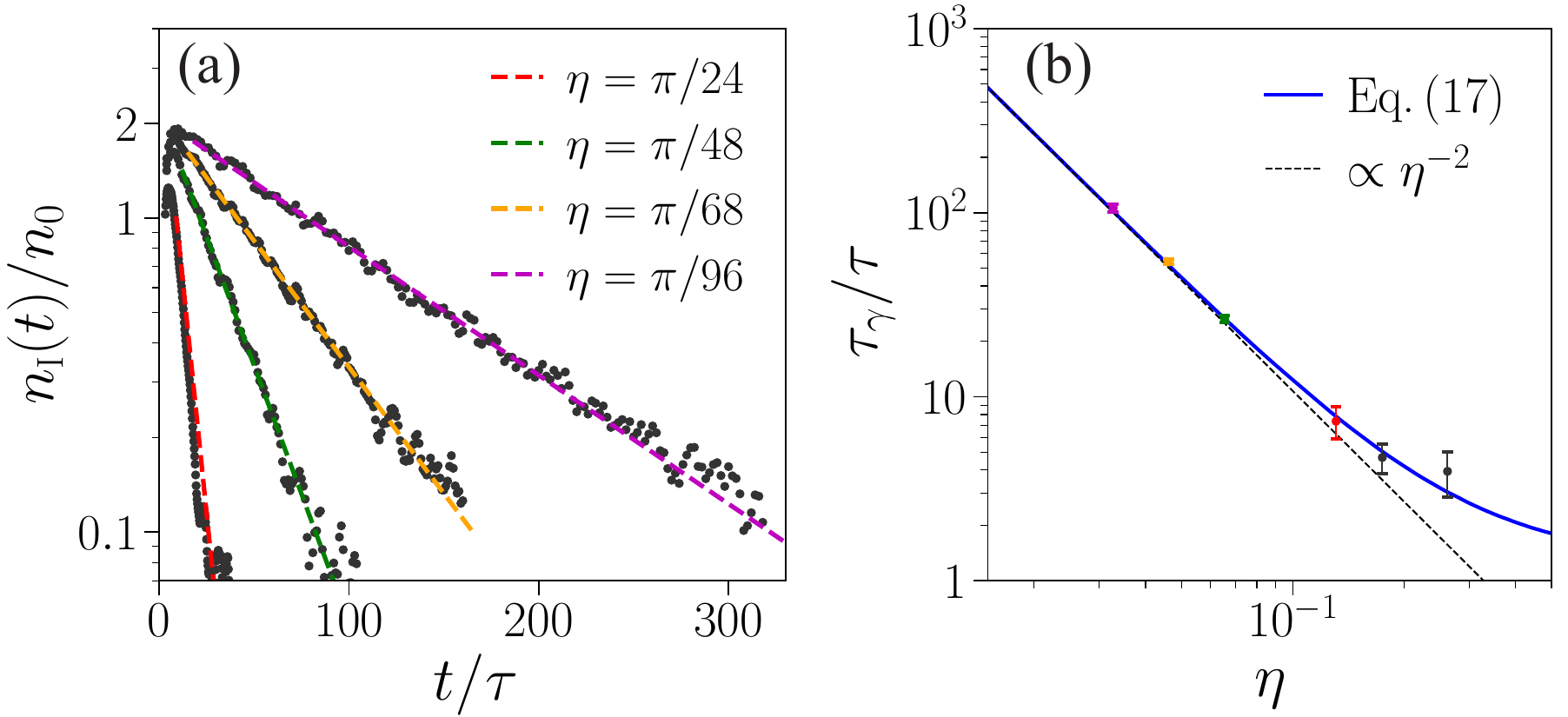}
    \caption{
     (a)~Semi-log plot of $n_{\mathrm{I}}(t) / n_0$ (black circles), with $n_{\mathrm{I}}(t)=n_{\mathrm{I}}(-\bm{k}_0\!-\!2\bm{\kappa}_x,t)$ obtained from the simulation, and $n_0  = 1 / (\pi^2\rho^2\gamma_0)$, as a function of time for different $\eta=\pi/n$ ($n=24,48,68,96$).
    The solid lines are linear fits. 
     (b)~Comparison between the dephasing times (colored circles with error bars) at $\eta=\pi/n$ ($n=12,18,24, 48,68,96$) and the theoretical prediction~Eq.~(\ref{eq:lifetime}) (blue solid line). The predicted scaling $\tau_{\gamma}\propto\eta^{-2}$ (black dashed line) holds for small $\eta$.
    All simulations have been performed with $\bm{k}_0=4\bm{\kappa}_x$.
    } 
    \label{fig:lifetime}
\end{center}
\end{figure}

For times $t\gg \tau$, we find that 
\begin{equation}\label{eq:nC_long_time}
    n^{\mathrm{C}}(\bm{k},t) = 2n_0 \,e^{-Dt\,(\bm{k} + \bm{k_0} +  2\bm{\kappa}_x)^2} \, e^{-t/\tau_\gamma},
\end{equation}
where $n_0  = 1 / (\pi^2\rho^2\gamma_0)$ is the value of $n(\bm{k},t)$ in the very long time diffusive limit, i.e., after full equilibration of the momentum distribution over the two branches of the Fermi surface.
The first exponential factor describes the diffusive dynamics that determine the scaling of the width of the CBS peak.
This term is similar to that found for spinless particles~\cite{Akkermans_2007_book}, except that the peak is now offset as already explained above. 
The second exponential factor shows that the gauge field causes the CBS peak to decay exponentially with a lifetime given by the dephasing time $\tau_\gamma$. 

In Fig.~\ref{fig:lifetime}(a), we show a semi-log plot of $n_{\mathrm{I}}(t) / n_0$, with $n_{\mathrm{I}}(t)=n_{\mathrm{I}}(-\bm{k}_0\!-\!2\bm{\kappa}_x,t)$, versus time obtained in the simulations.
Linear fits confirm an exponential decay, and their slopes give numerical estimates of the dephasing time.
In Fig.~\ref{fig:lifetime}(b) we compare these with the theoretical prediction Eq.~(\ref{eq:lifetime}).
The agreement is within the error bars in the simulation results and the expected scaling for small 
$\eta$ is recovered.

Letting $\eta\to 0$ at fixed disorder (constant $\ell$) and fixed gauge field strength $\kappa$, the gauge field $\hat{\bm{A}}$ becomes aligned with the $Ox$-axis. At $\eta=0$, the gauge field is fully removed by the non-Abelian gauge transformation to the new frame, see Eqs.~(\ref{eq:transformed_Ham}) and (\ref{eq:A_xy_primed}).
Since the random potential is spin-independent, the system then breaks up into two independent uncoupled subsystems with  orthogonal symmetry. 
The dephasing rate, described by Eq.~(\ref{eq:lifetime}), goes to zero (as seen in Fig.~\ref{fig:lifetime}) and a robust, i.e., non-transient, CBS peak with a finite angular offset is observed with respect to the original frame. In the new frame there is no offset (see Fig.~\ref{fig:shift_schematic} with $\kappa_x \to \kappa$).
The CBS dip can no longer be observed since, as there is no scattering between the $(\pm)$ branches of the Fermi surface, the occupation of the $(-)$ branch remains zero, see Fig.~\ref{fig:moment_distrib}(d).

The crossover, at fixed disorder, to the orthogonal class can also be obtained by letting $\kappa\to 0$ at fixed $\eta$, that is by letting the gauge field go to zero irrespective of its direction. 
This time the dephasing rate in Eq.~(\ref{eq:lifetime}) reduces to the known scaling $\tau^{-1}_{\gamma} \sim (\kappa\ell)^2 \, \tau^{-1}$~\cite{Dyakonov_1972,Wenk_2010}, associated with the Dyakonov--Perel spin relaxation mechanism (obtained in the limit $\kappa\ell \ll 1$).

{\it Discussion---}We studied the propagation of a spin-$1/2$ particle in a two-dimensional (2D) disordered spin-independent potential and subject to a general uniform SU(2) SOC.
We found that a transient CBS peak appears in the disorder-averaged momentum distribution
but with a momentum that is offset from the exact backscattering direction. 
The offset arises from momentum-dependent spin rotations induced by the non-Abelian gauge field, which result in a spin-state mismatch accumulated along a scattering path and its reciprocal partner, thereby displacing the interference maximum.
We presented a non-Abelian gauge transformation that allows an intuitive understanding of the offset.
We found that the time dependence of the peak shape is  well described by a theoretical calculation
of the Cooperon.
We identified a dephasing time $\tau_\gamma$ that describes the exponential decay of the Cooperon contribution to the
momentum distribution at times much longer than scattering mean free time $\tau$. We found excellent agreement between 
estimates of $\tau_\gamma$ extracted from the simulations and the analytic calculation.

A future direction is to explore the Anderson transition (AT)~\cite{Anderson_1958} that occurs 
in 2D with SOC at strong disorder~\cite{Evangelou_1987,T-Ando_1989, Asada_2002}. 
Changes in the time dependence of the CBS width identify the AT and 
the critical exponent extracted from the scaling of the width~\cite{Arabahmadi_2024} is consistent with that for the 2D symplectic universality class~\cite{Asada_2002}.
In the localized phase, a coherent forward scattering (CFS) peak emerges, which also characterizes the AT~\cite{Karpiuk_2012,Micklitz_2014,KL-Lee_2014b,Ghosh_2014,Ghosh_2017,Lemarie_2017,Hainaut_2018,Arrouas_2025}.

Cold atoms, where synthetic gauge fields have already been implemented~\cite{YJ-Lin_2011,P-Wang_2012,*Cheuk_2012,Huang_2016,Z-Wu_2016,Leroux_2018,ZY-Wang_2021,Q-Liang_2024,Madasu_2025}, are a promising platform for the experimental observation of the effects studied here.
Recently, the momentum-space dynamics of a wave packet subjected to a uniform synthetic SU(2) gauge field have been studied in~\cite{Hasan_2022} in a system described by an SOC Hamiltonian equivalent to ours~\cite{SM}. 
In experiments, the initial wave number dispersion $\Delta k_0$ induces a time decay of the CBS peak on a timescale $\tau_{\Delta}=[2D(\Delta k_0)^2]^{-1}$ even without SOC~\cite{Cherroret_2012,Jendrzejewski_2012a,Labeyrie_2012}. 
We expect the effective decay time $\tau_{\mathrm{eff}}$ of the CBS peak to be $1/\tau_{\mathrm{eff}} = 1/\tau_\gamma + 1/\tau_{\Delta}$. To resolve $\tau_\gamma$, we need $1/\tau_\Delta \lesssim 1/\tau_\gamma$. Using the small $\eta$ scaling for $\tau_\gamma$, this corresponds to $\Delta k_0 \ell \lesssim \eta$.

Finally, we compare electronic systems and cold atom systems.
One important difference is the timescale. In cold atom systems, the scattering time $\tau$ can be tuned to the millisecond range~\cite{Richard_2019}, making it easy to observe dynamics on timescales comparable to $\tau$. In contrast, in electronic systems such as quantum wells, $\tau$ is typically on the order of picoseconds~\cite{Tschirky_2017}, making it challenging to access dynamics on timescales comparable to $\tau$.
A method to observe weak-localization echoes using terahertz pulses has been theoretically proposed~\cite{ZL-Li_2024} as an analogue of cold atom experiments~\cite{Micklitz_2015,Muller_2015}. This approach would allow fast dynamics to be probed in electronic systems.
Tuning of $\eta$ can be achieved straightforwardly in electronic systems because the Rashba SOC is tunable~\cite{Nitta_1997}, and the regime corresponding to $\eta\approx0$ has been realized, e.g., in quantum wells~\cite{Koralek_2009}.

\begin{acknowledgments}
    {\it Acknowledgments---}This work was supported by JST SPRING (Grant No.~JPMJSP2138), by JSPS KAKENHI (Grant No.~JP25KJ1758), and by the French government, through the UCA$^\text{JEDI}$ ``Investissements d'Avenir" project, managed by the National Research Agency (ANR) (Grant No.~ANR-15-IDEX-01). 
    M.K. thanks the Program for Leading Graduate Schools: ``Interactive Materials Science Cadet Program" and the Research Fellowship for Young Scientists (Grant No.~JP25KJ1758) for support. M.K. also thanks UMR 7010 INPHYNI (UniCA and CNRS) for support and kind hospitality.
\end{acknowledgments}

\bibliography{refs_Letter}

\clearpage

\appendix
\onecolumngrid
\renewcommand{\thefigure}{S\number\numexpr\value{figure}-5\relax}
\renewcommand{\thetable}{S\Roman{table}}

\section{\large{Supplemental Material: Coherent transport in two-dimensional disordered potentials under spatially uniform SU(2) gauge fields}}

\section{A: Minimal 2D Hamiltonian equivalent to arbitrary spatially uniform SU(2) vector potentials}

We consider the following Hamiltonian in momentum space describing the dynamics of a spin-$\tfrac12$ particle subjected to a spatially uniform SU(2) non-Abelian vector potential in a two-dimensional (2D) space:
\begin{equation}\label{eq_SM:original_Ham}
    H_0(\bm{k}) = \frac{(\hbar\bm{k}\mathbbm{1}_2+\bm{A})^2}{2m} = \frac{\hbar^2\bm{k}^2}{2m}\mathbbm{1}_2 + \frac{\hbar\bm{k}\cdot\!\bm{A}}{m} + \frac{\bm{A}^2}{2m},
\end{equation}
where we have introduced the momentum of the particle under the form $\bm{p}=\hbar\bm{k}$, where $\bm{k}$ is the wave vector. Each space component of the vector potential is a linear combination of the Pauli spin matrices $\sigma_i$ ($i=1,2,3$) 
$$\bm{A}=M\bm{\sigma}$$ with
$M$ a $2\times 3$ real matrix and $\bm{\sigma}=(\sigma_1,\sigma_2,\sigma_3)^\top$ ($\top$ denotes transposition). Using  $$\sigma_a\sigma_b=\delta_{ab}\mathbbm{1}_2+i\sum_c\epsilon_{abc}\,\sigma_c,$$ where $\epsilon_{abc}$ is the Levi-Civita symbol, it can be shown that
\begin{equation}\label{A_squared}
    \bm{A}^2=(\sum_{i,j} M^2_{ij}) \, \mathbbm{1}_2 = \Tr(M M^\top )\mathbbm{1}_2 := \hbar^2 \kappa^2 \mathbbm{1}_2.
\end{equation}
This defines a momentum scale $\hbar \kappa$ and a corresponding energy scale
\begin{equation}
    E_\kappa=\frac{\hbar^2 \kappa^2}{2m}.
\end{equation}
Since the final term in the Hamiltonian is $\bm{k}$-independent and proportional to $\mathbbm{1}_2$, cf. Eq. (\ref{A_squared}), it can be removed by a redefinition of the origin of energy. 
Making use of a singular value decomposition (SVD) we now write $M$ in the form
\begin{equation}\label{eq_SM:SVD}
    M = \hbar \kappa\, R_1\Lambda R_2^\top,
\end{equation}
with $R_1\in\textrm{O}(2)$, $R_2\in\textrm{O}(3)$, and a $2\times 3$ real matrix $\Lambda$ of the form $$
\left(
\begin{matrix}
    \lambda_1 & 0 & 0 \\
    0 & \lambda_2 & 0
\end{matrix}\right),$$
with singular values $\lambda_1\ge0$ and $\lambda_2\ge0$.
While the SVD is not unique, if we order the singular values so that $\lambda_1\ge\lambda_2$, the matrix $\Lambda$ 
becomes unique.
It follows from Eq (\ref{A_squared}) that $\lambda_1^2+\lambda_2^2=1$. Thus we may write
\begin{equation}\label{eq_SM:singular_vals_2D}
    \lambda_1 = \cos\eta,\ \ \ 
    \lambda_2 = \sin\eta,
\end{equation}
We now define
\begin{equation}\label{eq_SM:rotation_momentum_spin} 
     \bm{r}' = R_1^\top  \bm{r}, \ \ \ \bm{k}' = R_1^\top\bm{k},\ \ \ \bm{\sigma}' = R_2^\top \bm{\sigma}.
\end{equation}
The Hamiltonian~(\ref{eq_SM:original_Ham}) can then be rewritten as
\begin{equation}\label{eq_SM:equivalent_Ham}
    H_0(\bm{k}') = \frac{\hbar^2\bm{k}'^{\,2}}{2m}\mathbbm{1}_2 
    + \hbar^2 \kappa \frac{\bm{k}'\!\cdot\!(\Lambda\bm{\sigma}')}{m} 
    + E_\kappa \mathbbm{1}_2.
\end{equation}
To ensure that the Pauli commutation relations are preserved, i.e., that we have $$[\sigma_a',\sigma_b']=2i \sum_c \epsilon_{abc}\, \sigma_c',$$
we need $R_2\in \textrm{SO}(3)$. Therefore, if $\det R_2=-1$, we make the replacements $R_1\to -R_1$ and $R_2\to -R_2$.
Next, if now $\det R_1=-1$, we make the further replacements
\begin{equation}
    R_1 \to R_1\! \left(\begin{matrix} 
        1 & 0  \\
        0 & -1  \\
    \end{matrix}\right)
    , \ \ 
     R_2 \to R_2 \!\left(\begin{matrix} 
        1 & 0  & 0 \\
        0 & -1 & 0\\
        0 & 0 & -1 \\
    \end{matrix}\right).
\end{equation}
After these replacements both $\det R_1=1$ and $\det R_2=1$, and we can express $M$ in the form
\begin{equation}
    M = \hbar \kappa R_1 
    \left(\begin{matrix}
       \cos \eta  & 0 & 0 \\
        0 & \sin \eta & 0
    \end{matrix}\right)
     R_2^\top,
\end{equation}
with between $\eta \in[0,\pi/4]$.
Dropping the primes, Hamiltonian~(\ref{eq_SM:original_Ham}) is reduced to
\begin{equation}\label{eq_SM:equivalent_Ham_2D}
  H_0(\bm{k})=\frac{\hbar^2(k_x^2+k_y^2)}{2m}\mathbbm{1}_2 
  + \hbar^2\kappa \frac{ k_x \cos \eta \\\ \sigma_1 + k_y \sin \eta \\\ \sigma_2}{m} 
  + E_{\kappa} \mathbbm{1}_2.
\end{equation}
We see that, after suitable transformations, two parameters, a momentum scale $\hbar \kappa$ and an angle $\eta$, are sufficient to characterize a spatially uniform SU(2) gauge field in 2D.

\subsection{\it Example 1: Rashba and 2D Dresselhaus spin--orbit couplings}
Hamiltonians $H_{\rm R}$ and $H_{\rm D}$ for Rashba and 2D Dresselhaus spin--orbit couplings, respectively,  have the forms~\cite{Winkler_2004}
\begin{equation}
    H_{\rm R}(\bm{k})=\frac{\hbar\alpha_{\rm R}}{m} \left( k_y \sigma_x - k_x \sigma_y\right),
\end{equation}
\begin{equation}
    H_{\rm D}(\bm{k})=\frac{\hbar\alpha_{\rm D}}{m} \left( k_y \sigma_y - k_x \sigma_x\right).
\end{equation}
with $\alpha_{\rm R}$ and $\alpha_{\rm D}$ real constants. For Rashba spin--orbit coupling we have
\begin{equation}
    M =  \alpha_{\rm R} \left(\begin{matrix}
        0 & -1 & 0 \\
        1 &  0 & 0 \\
    \end{matrix}\right)
    = \hbar \kappa
\left(\begin{matrix}
        1 & 0  \\
        0 & 1  \\
    \end{matrix}\right)
    \left(\begin{matrix}
        1 & 0 & 0 \\
        0 & 1 & 0 \\
    \end{matrix}\right)
        \left(\begin{matrix}
        0 & -1 & 0 \\
        1 &  0 & 0 \\
        0 &  0 & 1 \\
    \end{matrix}\right).
\end{equation}
This corresponds to $\hbar \kappa = \sqrt{2} \alpha_{\rm R}$ and $\eta=\pi/4$.
For Dresselhaus spin--orbit coupling we have
\begin{equation}
    M = \alpha_{\rm D} \left(\begin{matrix}
        -1 & 0 & 0 \\
         0 & 1 & 0 \\
    \end{matrix}\right)
    =  \hbar \kappa  
\left(\begin{matrix}
        1 & 0  \\
        0 & 1  \\
    \end{matrix}\right)
    \left(\begin{matrix}
        1 & 0 & 0 \\
        0 & 1 & 0 \\
    \end{matrix}\right)
        \left(\begin{matrix}
        -1 & 0 & 0 \\
        0 &  1 & 0 \\
        0 &  0 & -1 \\
    \end{matrix}\right).
\end{equation}
This corresponds to $\hbar \kappa = \sqrt{2} \alpha_{\rm D}$ and $\eta=\pi/4$.
For a mixture of Rashba and Dresselhaus spin--orbit couplings
\begin{equation}
    H_{\mathrm{SO}} =  H_{\rm R} + H_{\rm D}.
\end{equation}
For $\alpha_{\rm R} \ge \alpha_{\rm D} \ge 0$
\begin{equation}
    M = \left(\begin{matrix}
        -\alpha_{\rm D} & -\alpha_{\rm R} & 0 \\
         \alpha_{\rm R} &  \alpha_{\rm D} & 0 \\
    \end{matrix}\right)
    = \hbar \kappa
    \frac{1}{\sqrt{2}} 
\left(\begin{matrix} 
        1 & 1  \\
        -1 & 1  \\
    \end{matrix}\right)
    \left(\begin{matrix}
        \lambda_1  & 0 & 0 \\
        0 &  \lambda_2 & 0 \\
    \end{matrix}\right)
        \frac{1}{\sqrt{2}}
        \left(\begin{matrix}
        -1 &  1 &  0 \\
        -1 & -1 &  0 \\
         0 &  0 & \sqrt{2} \\
    \end{matrix}\right)
\end{equation}
and for $\alpha_{\rm D} \ge \alpha_{\rm R}  \ge0$
\begin{equation}
    M = \left(\begin{matrix}
        -\alpha_{\rm D} & -\alpha_{\rm R} & 0 \\
         \alpha_{\rm R} &  \alpha_{\rm D} & 0 \\
    \end{matrix}\right)
    = 
   \hbar \kappa
   \frac{1}{\sqrt{2}} 
\left(\begin{matrix} 
        1 & 1  \\
        -1 & 1  \\
    \end{matrix}\right)
    \left(\begin{matrix}
        \lambda_1  & 0 & 0 \\
        0 &  \lambda_2 & 0 \\
    \end{matrix}\right)
        \frac{1}{\sqrt{2}}
        \left(\begin{matrix}
        -1 & -1 &  0 \\
        -1 &  1 &  0 \\
         0 &  0 & -\sqrt{2} \\
    \end{matrix}\right)
\end{equation}
with 
\begin{equation}
    \lambda_1 = \frac{1}{\sqrt{2}} \frac{\alpha_{\rm R}+\alpha_{\rm D}}{\sqrt{\alpha_{\rm R}^2+\alpha_{\rm D}^2}} ,\ \ \ 
    \lambda_2 = \frac{1}{\sqrt{2}} \frac{\left|\alpha_{\rm R}-\alpha_{\rm D}\right|}{\sqrt{\alpha_{\rm R}^2+\alpha_{\rm D}^2}} ,   \ \ \
    \hbar\kappa = \sqrt{2\left(\alpha_{\rm R}^2+\alpha_{\rm D}^2\right)} ,\ \ \
    \eta = \arctan\left( \frac{\left|\alpha_{\rm R}-\alpha_{\rm D}\right|}{\alpha_{\rm R} + \alpha_{\rm D}}\right).
\end{equation}
The special case $\alpha_{\rm D} = \alpha_{\rm R}$ is equivalent to Ref.~\cite{Bernevig_2006} and corresponds to $\eta=0$.

\subsection{\it Example 2: Cold atomic synthetic non-Abelian gauge fields}

Hamiltonian Eq.~(\ref{eq_SM:equivalent_Ham_2D}) can also be realized in systems of ultracold atoms.
While several strategies exist for implementing synthetic non-Abelian gauge fields in ultracold atoms, we focus here on the approach based on degenerate dark states~\cite{Ruseckas_2005,Leroux_2018,Hasan_2022} where atoms are illuminated by suitable monochromatic laser beams. In this context, the scalar potential $\bm{A}^2/2m$ is no longer proportional to $\mathbbm{1}_2$ but can nevertheless be removed by appropriately tuning each laser frequency off resonance from the atomic transition. The effective Hamiltonian for the dark-state manifold then reads
\begin{equation}\label{eq_SM:effective_Ham_synthetic_SOC}
    H_{\rm eff}(\bm{k}) = \frac{\big(\hbar\bm{k}\mathbbm{1}_2+\bm{A}\big)^2}{2m}.
\end{equation}
We consider the laser tripod configuration that was used in Ref.~\cite{Hasan_2022} where the laser beam $i$ addresses the atomic transition $\ket{i}\leftrightarrow\ket{4}$ ($i=1,2,3$). The corresponding space-dependent Rabi frequencies are $\Omega_1 \, e^{ik_{\rm L}x}$, $\Omega_2 \, e^{ik_{\rm L}y}$, and $\Omega_3 \, e^{-ik_{\rm L}x}$, $k_{\rm L}$ being the laser wavenumber. We further parametrize the Rabi frequencies using spherical coordinates by writing $\bm{\Omega} = (\Omega_1, \Omega_2, \Omega_3) = \Omega_0 \, \tilde{\bm{\Omega}}$, where $\Omega^2_0=\sqrt{\Omega^2_1+ \Omega^2_2+\Omega^2_3}$ and $\tilde{\bm{\Omega}} = (\cos\varphi\sin\vartheta,\sin\varphi\sin\vartheta,\cos\vartheta)$.  
Following Ref.~\cite{Ruseckas_2005}, The dark states corresponding to this laser scheme are 
\begin{equation}
\begin{aligned}
    \ket{D_1(k_{\rm L},\varphi,\vartheta)} &= e^{-2ik_{\rm L}x}\sin\varphi\,\ket{1} - e^{-ik_{\rm L}(x+y)}\cos\varphi\,\ket{2},\\
    \ket{D_2(k_{\rm L},\varphi,\vartheta)} &= e^{-2ik_{\rm L}x}\cos\varphi\cos\vartheta\,\ket{1} + e^{-ik_{\rm L}(x+y)}\sin\varphi\cos\vartheta\,\ket{2} - \sin\vartheta\,\ket{3}.
\end{aligned}
\end{equation}
The vector potential $\bm{A}_{ij}=-i\hbar\braket{D_i|\nabla D_j}$ is then written in the form $\bm{A}=M\bm{\sigma}+\bm{c}$ with
\begin{equation}
\begin{aligned}\label{eq_SM:M_and_c}
M = \frac{\hbar k_{\rm L}}{2}\left(
\begin{matrix}
     -2\sin\varphi\cos\varphi\cos\vartheta & 0 
    & 1 -2\sin^2\varphi-2\sin^2\vartheta + \sin^2\varphi\sin^2\vartheta \\
    2\sin\varphi\cos\varphi\cos\vartheta & 0 
    & -\cos^2\varphi+\sin^2\varphi\cos^2\vartheta
\end{matrix}
\right),\quad
\bm{c} = \frac{\hbar k_{\rm L}}{2}\left(
\begin{matrix} 
    -3 + 2\sin^2\vartheta - \sin^2\varphi\sin^2\vartheta \\
    -1+\sin^2\varphi\sin^2\vartheta
\end{matrix}
\right).
\end{aligned}
\end{equation}
The constant vector $\bm{c}$ can be eliminated by shifting momenta, $\hbar\bm{k}\to\hbar\bm{k}+\bm{c}$.
Suitable rotations further reduce the vector potential to $\bm{A}'=\hbar \kappa \, (\cos\eta \,\sigma_1, \sin\eta \,\sigma_2)$ for any values of $k_{\rm L}$, $\varphi$ and $\vartheta$. A singular value decomposition of the matrix $M$ in Eq.~(\ref{eq_SM:M_and_c}) determines 
$\eta$ as a function of $\varphi$ and $\vartheta$. Though their analytical expressions are too involved, and not very illuminating, to be shown here, it shows that $\eta$ can be continuously varied by tuning the laser field strengths as long as the atomic evolution remains restricted to the dark-state manifold.
Equal Rabi frequency amplitudes used in Ref.~\cite{Hasan_2022} yields $\kappa = 2 k_{\textrm{L}}/3$ and $\eta = \pi/6$.
The special condition $(\Omega_1,\Omega_2,\Omega_3) = \Omega_0(\sqrt{a},a,\sqrt{a})$ with $a=\sqrt{2}-1$ yields $\eta=\pi/4$, whereas $\varphi=0,\pi/2$ or $\vartheta=0,\pi/2$ gives $\eta=0$. The latter corresponds to the case where at least one of the Rabi frequencies, $\Omega_1$, $\Omega_2$, or $\Omega_3$, is zero.

\subsection{B: Minimal 3D Hamiltonian with uniform SU(2) vector potentials}

The  discussion above can be extended to spatial dimension $d=3$. The wave vector and the vector potential become $\bm{k}=(k_x,k_y,k_z)$ and $\bm{A}=M\bm{\sigma}$, respectively, where $M$ is a ${3\times3}$ real matrix.
First, as for $d=2$, we use an SVD to express $M$ in the form
\begin{equation}
    M = \hbar\kappa\, R_3\Lambda R_4^\top,
\end{equation}
where $R_3, R_4 \in \textrm{O}(3)$, and $\Lambda=\textrm{diag}(\lambda_1,\lambda_2,\lambda_3)$ with $\lambda_1\ge\lambda_2\ge\lambda_3\ge0$.
To preserve the commutation relations for both orbital and spin angular momenta operators, we require $R_3\in\textrm{SO}(3)$ and $R_4\in\textrm{SO}(3)$. 
If $\det R_4=-1$, we make the replacements $R_3\to -R_3$ and $R_4\to -R_4$.
Next, if now $\det R_3=-1$, we make the further replacements
\begin{equation}
    R_3 \to R_3\! \left(\begin{matrix} 
        1 & 0  & 0 \\
        0 & 1 & 0\\
        0 & 0 & -1 \\
    \end{matrix}\right)
    , \ \
     \Lambda \to \left(\begin{matrix} 
        1 & 0  & 0 \\
        0 & 1 & 0\\
        0 & 0 & -1 \\
    \end{matrix}\right)\! \Lambda.
\end{equation}
After these replacements both $\det R_3=1$ and $\det R_4=1$, and we can express $M$ in the form
\begin{equation} 
    M = \hbar\kappa\, R_3
    \left( 
    \begin{matrix} 
    \lambda_1 & 0 & 0 \\ 
    0 &\lambda_2 & 0 \\ 
    0 & 0 & (\det M)\,\lambda_3 
    \end{matrix} 
    \right) R_4^\top. 
\end{equation}
After the transformations 
$$\bm{r}' = R_3^\top\bm{r},\ \ \bm{k}' = R_3^\top\bm{k}, \ \ \textrm{and}\ \ \bm{\sigma}' = R_4^\top\bm{\sigma},$$
the gauge field in the new frame consists of three spatial components, proportional to $\sigma_1$, $\sigma_2$, and $\sigma_3$, respectively.

\subsection{C: Eigenstates, eigenenergies and Berry phase for the clean Hamiltonian}

\noindent {\it Diagonalization}---In the full Hilbert space,  the clean Hamiltonian in Eq.~(\ref{eq_SM:equivalent_Ham_2D}) reads
\begin{equation}
    \hat{H}_0 = \sum_{\sigma,\sigma'=\uparrow,\downarrow} \int\! \frac{d^2\bm{k}}{(2\pi)^2}\,  H^{\sigma\sigma'}_0(\bm{k}) \, \ket{\bm{k},\sigma}\!\bra{\bm{k},\sigma'} = \sum_{s=\pm} \int\! \frac{d^2\bm{k}}{(2\pi)^2}\, E_s(\bm{k}) \, \ket{\psi_s(\bm{k})}\!\bra{\psi_s(\bm{k})},
\end{equation}
where $\ket{\sigma=\uparrow, \downarrow}$ the spin states along the $z$ axis and the matrix entries in spin space are given by $H^{\sigma\sigma'}_0(\bm{k}) = \bra{\sigma} H_0(\bm{k}) \ket{\sigma'}$. Using the polar coordinates of $\bm{k}$ ($k_x = k\cos\theta$, $k_y = k\sin\theta$), the full eigenstates of $\hat{H}_0$ are $|\psi_{s}(\bm{k})\rangle = |\bm{k}\rangle\otimes|S_{s}(\theta)\rangle$, where $s=\pm$ denotes the branch index. We have
\begin{equation}
    \label{eq_SM:spinor}
    |S_{+}(\theta)\rangle = \frac{|\!\uparrow\rangle + e^{i\phi_{\eta}(\theta)}|\!\downarrow\rangle}{\sqrt{2}}, 
    \hspace{8mm}
    |S_{-}(\theta)\rangle = \frac{-e^{-i\phi_{\eta}(\theta)}|\!\uparrow\rangle +|\!\downarrow\rangle}{\sqrt{2}},
\end{equation}
where $\phi_{\eta}(\theta)$ is given by
\begin{align}\label{eq_SM:def_phi_eta}
    \phi_{\eta}(\theta) = \arg(\cos\eta\cos\theta + i\sin\eta\sin\theta).
\end{align}
The corresponding eigenenergies are 
\begin{equation}
\label{eq_SM:eigenE}
    E_{\pm}(\bm{k}) = \frac{\hbar^2(k_x^2+k_y^2)}{2m} \pm \frac{\hbar^2\kappa }{m} \sqrt{ k_x^2\cos^2\!\eta +k_y^2\sin^2\!\eta} + E_\kappa = \frac{\hbar^2k^2}{2m} \pm \frac{\hbar^2\kappa k}{m} \sqrt{\frac{1 + \cos2\eta\cos2\theta}{2}} + E_\kappa.
\end{equation}
\\
\noindent {\it Fermi surface---}The $(+)$ branch and $(-)$ branch of the Fermi surface are obtained by slicing $E_{\pm}(\bm{k}) $ at the considered positive constant initial energy $E_{0}= E_{+}(\bm{k}_0)$. Solving for $E_{\pm}(\bm{k}) = E_{0}$, we get
\begin{equation}
    \frac{k_\pm(\theta)}{\kappa} 
    = \sqrt{(\varepsilon_0\!-\!1) + \frac{1+\cos2\eta \cos2\theta}{2}} 
    \mp \sqrt{\frac{1+\cos2\eta \cos2\theta}{2}} 
    = \sqrt{\frac{2\varepsilon_0 \!-\!1+\cos2\eta \cos2\theta}{2}} \mp \sqrt{\frac{1+\cos2\eta \cos2\theta}{2}},
\end{equation}
where $\varepsilon_0 = E_{0}/E_{\kappa} = 2mE_{0}/\hbar\kappa^2$.
The separation between the two branches of the Fermi surface can be quantified by $\Delta k(\theta) = k_{-}(\theta)-k_{+}(\theta) = \kappa \sqrt{2+2\cos2\eta \cos2\theta}$. Since $0\le \eta\le \pi/4$, the two branches are closest at $\theta=\pi/2$ where their momentum separation is $\Delta k = 2\kappa|\sin\eta|$. When $\eta=0$, the two branches cross at $\theta=\pi/2$. When $\eta=\pi/4$, the ($+$) and ($-$) branches of the Fermi surface are two nested circles centered at the origin with respective radii $k_0$ and $k_0+\sqrt{2}\,\kappa$, differing by $\Delta k=\sqrt{2}\,\kappa$.\\

\noindent {\it Berry phase}---Following Ref.~\cite{Fuchs_2010}, we obtain the Berry connection $\bm{a}_{\pm}(\bm{k}) = i\langle S_{\pm}(\bm{k})|\nabla_{\!\bm{k}}S_{\pm}(\bm{k})\rangle$ associated with each spin eigenstate: 
\begin{equation}
    \bm{a}_{\pm}(\bm{k}) = \mp \frac{(\bm{\nabla}_{\!\bm{k}}\phi_\eta)}{2}=\mp\frac{1}{2k}\frac{\sin2\eta}{1+\cos2\eta\cos2\theta}\,\va{e}_{\theta}.
\end{equation}
The Berry curvature being $\bm{\Omega}_{\pm}(\bm{k}) = \nabla_{\!\bm{k}}\times\bm{a}_{\pm}(\bm{k})$, we see that it is zero as soon as $k\neq 0$ and we thus anticipate a singularity at the origin:
\begin{equation}
    \bm{\Omega}_{\pm}(\bm{k}) = \mp\, \Omega_\eta \,\delta (\bm{k}) \,\va{e}_{z}.
\end{equation}
To obtain $\Omega_\eta$, we compute the Berry phase $\gamma_\pm[C] = \oint_C \bm{a}_\pm(\bm{k})\cdot d\bm{k} = \mp\, \Omega_\eta$ associated to any closed path $C$ in momentum space around $\bm{k}=\bm{0}$. Using $\int_0^{2\pi} dx/(a+\cos x) = 2\pi/\sqrt{a^2-1}$ for $a >1$, we find:
\begin{equation}
    \Omega_\eta = \rm{sign}(\eta) \,\pi. 
\end{equation}
We note that a Berry phase can also be defined in the two-dimensional parameter space spanned by $\kappa$ and $\eta$ at fixed wave vector $\bm{k}$.
As is clear from  Eq.~(\ref{eq_SM:def_phi_eta}), $\eta$ and $\theta$ are interchangeable. Hence, the same result is obtained in the $(\kappa,\eta$) space. 
In other words, the Berry phase $\pm\pi$ is obtained for any closed path enclosing $\kappa=0$.

\subsection{D: Scattering properties of the spin-independent disordered potential}

\noindent {\it Transitions between clean eigenstates}---We consider a disorder potential of the form $$\hat{V}=\int\! d\bm{r} \, V(\bm{r}) \, \ket{\bm{r}}\bra{\bm{r}} \otimes \mathbbm{1}_{{\rm spin}},$$
that is diagonal in spin space. The matrix elements between eigenstates of the Hamiltonian of the clean system, i.e., the eigenstates of the Hamiltonian without the random potential, factor as follows 
\begin{equation}
    \bra{\psi_{s}(\bm{k})} \hat{V} \ket{\psi_{s'}(\bm{k'})} = V(\bm{k}-\bm{k}') \ \langle S_{s}(\theta)| S_{s'}(\theta')\rangle,
\end{equation}
where $V(\bm{k}) = \int d\bm{r} \, V(\bm{r}) \, \exp(-i \bm{k}\cdot\bm{r})$ is the Fourier transform of $V(\bm{r})$.
This shows that, even if the disorder potential is spin-independent, transition matrix elements between clean eigenstates are spin dependent. The disorder-induced transition probability between clean eigenstates contains a factor
\begin{equation}
    P_{ss'}(\theta,\theta', \eta) =
    |\langle S_{s}(\theta)|S_{s'}(\theta')\rangle|^2 = \frac{1+\cos\Delta\phi_\eta}{2}\, \delta_{ss'}
    + \frac{1-\cos\Delta\phi_\eta}{2} \, (1\!-\!\delta_{ss'}),
\end{equation}
where $\Delta\phi_\eta = \phi_\eta(\theta)-\phi_\eta(\theta')$. When $\eta = 0$, it follows from Eq.~(\ref{eq_SM:spinor}) that $\phi_{\eta}(\theta) = 0$ for $k_x > 0$ and $\phi_{\eta}(\theta) = \pi$ for $k_x < 0$, regardless of $k_y$.
In this case, $\cos\Delta\phi_{\eta=0} = {\rm sign}(k_xk'_x)$ and 
$$\langle\psi_s(\bm{k})|\hat{V}|\psi_{s}(\bm{k}')\rangle = 0 \hspace{0.25cm}  {\rm for} \ k_xk_x'<0 \hspace{0.5cm} {\rm and} \hspace{0.5cm} 
\langle\psi_s(\bm{k})|\hat{V}|\psi_{-s}(\bm{k}')\rangle = 0  \hspace{0.25cm} {\rm for} \ k_xk_x'>0.$$
Starting from an initial plane-wave state $\bm{k}_0$ located, for example, on the ($+$) branch with $\bm{k}_{0x} > 0$, this means that, when $\eta=0$, only the ($+$) branch with $k_x \ge 0$ and the ($-$) branch with $k_x \le 0$ are filled by successive scattering events. These two sectors of the two branches combine into a single circle centered at ($-\hbar\kappa,0$), as seen in Fig.~1(d) of the main text.\\

\noindent {\it Scattering mean free time and mean free path}---Following Ref.~\cite{Kuhn_2007}, the scattering mean free time $\tau(E)$ at energy $E$ is computed from the self-energy $\Sigma^R(E)$ through
\begin{equation}
\label{eq_SM:Scattime}
    \frac{\hbar}{2\tau(E)} = - {\rm Im}\,\Sigma^R(E).
\end{equation}
Within the Born approximation, 
$\Sigma^R(E) \approx \overline{\hat{V}G^R_0(E)\hat{V}}$ where $G^R_0(E) = (E-\hat{H}_0 +i0^+)^{-1}$ and $\overline{(\cdot\cdot\cdot)}$ denotes disorder average. A simple calculation provides
\begin{equation}
    \bra{{\bm k}}\Sigma^R(E) \ket{\bm{k}'} = \Sigma^R(\bm{k},\bm{k}',E) \simeq \int\!\!\frac{d^2\bm{q}}{(2\pi)^2} \overline{V(\bm{k}\!-\!\bm{q})V(\bm{q}\!-\!\bm{k}')}G_0^{R}(\bm{q},E).
\end{equation}
Here $G_0^{R}(\bm{k},E) = (E-H_0(\bm{k}) +i0^+)^{-1}$ is the free retarded Green's function associated to the spin-space Hamiltonian Eq.~(\ref{eq_SM:equivalent_Ham_2D}). In the $\ket{\uparrow,\downarrow}$ basis, we have
\begin{align}
\label{eq_SM:G0p}
    G_0^{R}(\bm{k},E)
    = \frac{g_{0+}^{R}(\bm{k},E) + g_{0-}^{R}(\bm{k},E)}{2} \mathbbm{1}_2
    + \frac{g_{0+}^{R}(\bm{k},E) - g_{0-}^{R}(\bm{k},E)}{2}
    \left[\cos\phi_{\eta}(\theta) \, \sigma_1 + \sin\phi_{\eta}(\theta) \, \sigma_2\right],
\end{align}
where
\begin{align}
    g_{0\pm}^{R}(\bm{k},E) &= \frac{1}{E - E_{\pm}(\bm{k}) + i0^+}.
\end{align}
Assuming that disorder average restores translation invariance, the two-point potential correlator in real space only depends on spatial separations, namely $ \overline{V(\bm{r})V(\bm{r}')} = C(\bm{r}\!-\!\bm{r}')$. It easily follows that 
\begin{align}\label{eq_SM:tp_corr_momentum_space}
    \overline{V(\bm{k}\!-\!\bm{q})V(\bm{q}\!-\!\bm{k}')} = (2\pi)^2\,\delta(\bm{k}\!-\!\bm{k}') C(\bm{k}\!-\!\bm{q}),
\end{align}
where $C(\bm{k})$ is the Fourier transform of $C(\bm{r})$.
The self-energy $\Sigma^R(E)$ is thus diagonal in momentum space and we have $\bra{{\bm k}}\Sigma^R(E) \ket{\bm{k}'} = (2\pi)^2 \, \delta (\bm{k}-\bm{k}') \, \Sigma^R(\bm{k},E)$ with
\begin{equation}\label{eq_SM:self-energy}
    \Sigma^R(\bm{k},E) = \int\!\!\frac{d^2\bm{q}}{(2\pi)^2}C(\bm{k}\!-\!\bm{q})G_0^{R}(\bm{q},E).
\end{equation}
In the context of this paper, $C(\bm{r}) = \gamma_0 \, \delta(\bm{r})$, implying $C(\bm{k})=\gamma_0$, and $\Sigma^R(\bm{k},E)$ becomes $\bm{k}$-independent. From Eq.~(\ref{eq_SM:G0p}), and using $\phi_{\eta}(\theta+\pi)=\phi_{\eta}(\theta)+\pi$ and $g_{0\pm}^{R}(-\bm{k},E)= g_{0\pm}^{R}(\bm{k},E)$, we find
\begin{equation}
  \Sigma^R(\bm{k},E) = \frac12\gamma_0 \, \int\!\!\frac{d^2\bm{q}}{(2\pi)^2}\left(\sum_{s=\pm} g^R_{0s}(\bm{q},E)\right) \, \mathbbm{1}_2.
\end{equation}
Introducing the total clean density of states at energy $E$
\begin{equation}
    \rho(E) =
    -\frac{1}{\pi}\int\!\!\frac{d^2\bm{q}}{(2\pi)^2}\, \textrm{Im}\!\left[g_{0+}^{R}(\bm{q},E) + g_{0-}^{R}(\bm{q},E)\right],
\end{equation}
where ${\rm Im\,}g^R_{0s}(\bm{q},E) = -i \pi \delta\!\left(E-E_s(\bm{q})\right)$, we find the scattering mean free time
\begin{equation}
    \tau = \frac{\hbar}{\pi\rho\gamma_0}.
\end{equation}
where $\rho = \rho (E_{0})$ is computed at the initial energy. Here, the density of states (including both spin states) per unit area can be calculated as $\rho \simeq m/(\pi\hbar^2)$, giving $\tau \simeq \hbar^3/(m\gamma_0)$. The scattering mean free path at energy $E=E_{0}$ is then  
given by $\ell = v_g\tau$, where $\bm{v}_g = \left(\boldsymbol{\nabla}_{\bm{k}}E_{+}\right)(\bm{k}_0)$ is the initial group velocity.\\

\noindent {\it Disorder-averaged Green's function in the $\ket{\rightleftarrows}$ basis}---Followed by the basis transformation from the $\ket{\uparrow,\downarrow}$ basis to the $\ket{\rightleftarrows}$ basis
\begin{equation}
    \ket{\rightarrow} = \frac{\ket{\uparrow}+\ket{\downarrow}}{\sqrt{2}},
    \hspace{8mm} \ket{\leftarrow} = \frac{\ket{\uparrow}-\ket{\downarrow}}{\sqrt{2}},
\end{equation}
we obtain the disorder-averaged retarded Green's function in the $\ket{\rightleftarrows}$ basis as
\begin{align}
    \overline{G}\ \!^{\rightleftarrows}(\bm{k},E) &= \frac12 \left[
    \frac{\mathbbm{1}_2 + \cos\phi_\eta(\theta)\,\sigma_3 - \sin\phi_\eta(\theta)\,\sigma_2}{E-E_+(\bm{k})+i\frac{\hbar}{2\tau}}
    +\frac{\mathbbm{1}_2 - \cos\phi_\eta(\theta)\,\sigma_3 + \sin\phi_\eta(\theta)\,\sigma_2}{E-E_-(\bm{k})+i\frac{\hbar}{2\tau}}
    \right].
\end{align}
Specifically,
\begin{equation}\label{eq_SM:GF_rightleftarrows}
    \overline{G}_{\rightarrow}(\bm{k},E) = \bra{\rightarrow}\overline{G}\ \!^{\rightleftarrows}(\bm{k},E)\ket{\rightarrow}
    =\frac{\cos^2\!\big[\phi_\eta(\theta)/2\big]}{E-E_+(\bm{k})+ i\frac{\hbar}{2\tau}} + \frac{\sin^2\!\big[\phi_\eta(\theta)/2\big]}{E-E_-(\bm{k})+ i\frac{\hbar}{2\tau}},
\end{equation}
where $\phi_\eta(\theta)$ is given in Eq.~(\ref{eq_SM:def_phi_eta}).
When $\eta=0$, Eq.~(\ref{eq_SM:GF_rightleftarrows}) becomes
\begin{equation}
    \overline{G}_{\rightarrow}(\bm{k},E) = \Big[E-\frac{\hbar^2(\bm{k}\!+\!\bm{\kappa}_x)^2}{2m}+i\frac{\hbar}{2\tau}\Big]^{-1},
\end{equation}
because $\phi_{\eta}(\theta) = 0$ for $k_x > 0$ and $\phi_{\eta}(\theta) = \pi$ for $k_x < 0$, regardless of $k_y$.

\section{E: Scaling for the dephasing time}

At $\theta=\pi/2$, we have $\Delta k= k_{-}\!-k_{+} = 2\kappa \sin \eta$ and thus a corresponding energy scale $E_{\Delta k} := \hbar^2(\Delta k)^2/(2m)= 4E_\kappa \sin^2\eta$. Meanwhile $1/\ell$ sets a scale for momentum and thus a corresponding disorder energy scale $E_\ell = \hbar^2/(2m\ell^2) = E_\kappa/(\kappa\ell)^2$. Defining $\tau_\ell = \hbar/E_\ell$ and $\tau_{\Delta k}=\hbar/E_{\Delta k}$, we have $\tau_{\Delta k}/\tau_\ell = (2\kappa_y\ell)^{-2} \approx (2\kappa \ell \eta)^{-2}$ when $\eta$ is small as we have considered in the paper. This ratio of time scales  characterizes the transition between the two branches at $\theta=\pi/2$.

\section{F: Parameters for the numerical simulations}

In the numerical simulation, we model a $\delta$-correlated potential in continuous space by a spatially uncorrelated random potential placed on a discrete $N_x\times N_y$ lattice with spacing $a$.
The value on each site is independently drawn from a uniform distribution over the interval $[-W/2,W/2]$.
For this potential, the spatial correlation reads
\begin{equation}
    \overline{V(\bm{r}_i)V(\bm{r}_j)} = \frac{W^2a^2}{12}\frac{\delta_{ij}}{a^2},
\end{equation}
where $\delta_{ij}/a^2 \to \delta(\bm{r}_i-\bm{r}_j)$ when $a \to 0$. From the comparison with $\overline{V(\bm{r})V(\bm{r}')} = \gamma_0 \, \delta(\bm{r}-\bm{r}')$, we infer that for
\begin{equation}\label{eq_SM:corr_amp_uniform_distrib}
    \gamma_0 = \frac{W^2a^2}{12}
\end{equation}
to be kept constant as $a\to 0$, we should let $W\to \infty$ while keeping the product $Wa$ constant.
When the lattice spacing $a$ is sufficiently smaller than the mean free path $\ell$ and the wavelength $\lambda=1/k_0$, the discrete potential is a good approximation to a $\delta$-correlated potential in continuous space.
\begin{figure}[t]
\begin{center}
    \includegraphics[width=0.36\columnwidth]{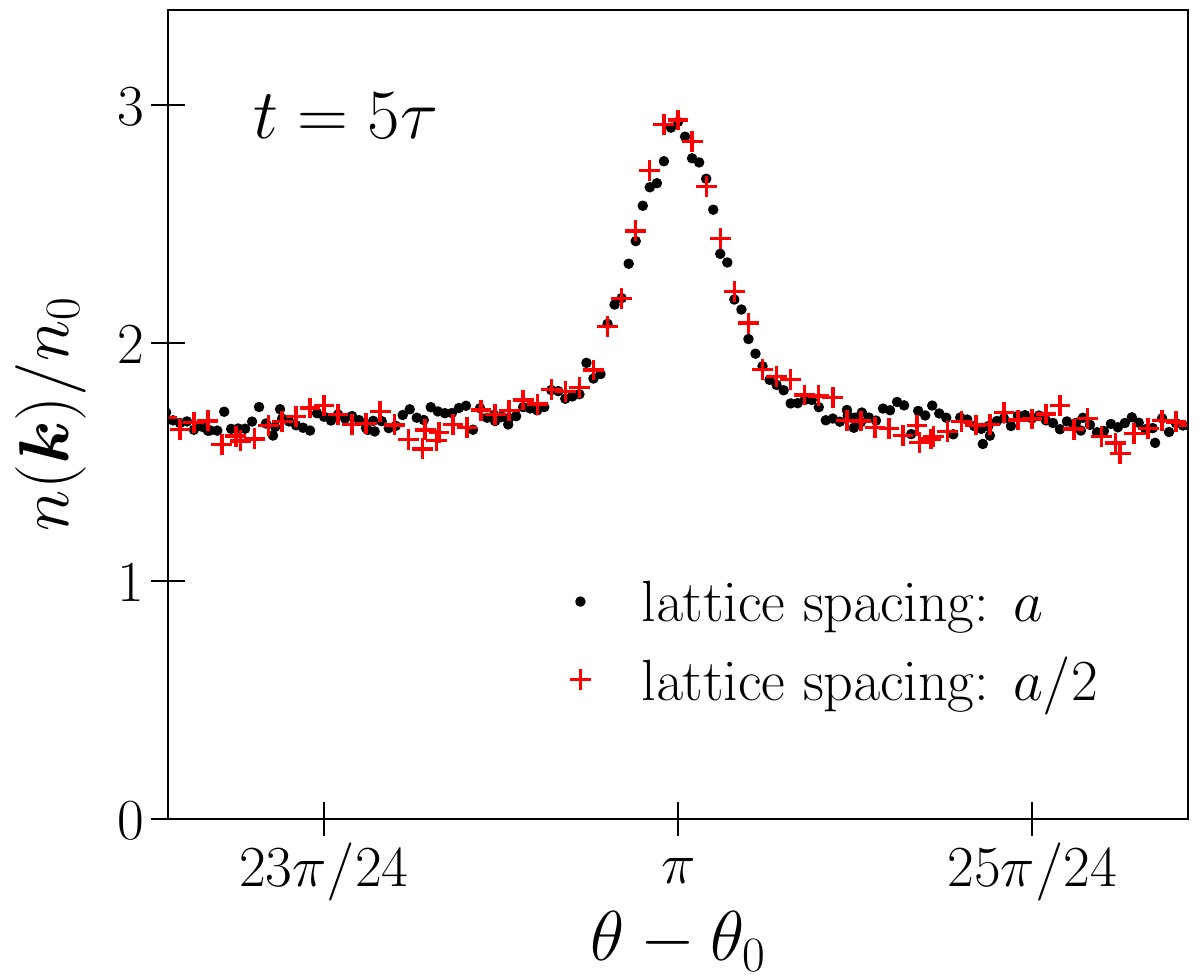}
    \caption{
    Comparison of momentum distributions when the lattice spacing $a$ given in Table~\ref{tab:simulation_params} is halved.
    $n_0  = 1 / (\pi^2\rho^2\gamma_0)$ is the value of $n(\bm{k},t)$ in the very long time diffusive limit.
    The quantitative agreement between the two distributions shows that the lattice spacing $a$ used in our numerical simulations is sufficiently small to avoid any mesh effects.
    } 
    \label{fig_SM:check_convergence}
\end{center}
\end{figure}

Table~\ref{tab:simulation_params} summarizes the parameters that we have used in our numerical simulations and Fig.~\ref{fig_SM:check_convergence} shows that the lattice spacing $a$ used in our simulations is small enough to obtain results insensitive to discretization effects.
We make a rough estimate of the mean free path $\ell$ for the initial wave vector $\bm{k}_0= 4\kappa\cos\eta\, \va{e}_{x}$ used in our simulations by taking $\cos\eta \approx 1$. This particular choice of the initial wave vector does not carry any particular physical significance. 
It is a numerical convenience intended to place the initial state and observed transient peak on the lattice grid. For this value, the initial group velocity is approximately $v_g = 10\,\hbar\kappa/m$. With our choice $\gamma_0 = 2\, E_{\kappa}^2 L_{\kappa}^2$, we find $\ell=v_g\tau\simeq 10\,L_\kappa $.
The dimensionless quantity 
$k_0\ell \simeq 40 \gg1$ which justifies the weak-disorder approximation used in the main text, i.e., treating the disordered potential perturbatively.
The numerical values of the scattering mean free paths associated to the other initial wave vector states that we have used in simulations do not differ significantly.

\renewcommand{\arraystretch}{1.8}
\setlength{\tabcolsep}{6pt}
\begin{table}[h]
\centering
\begin{tabular}{c}
\begin{tabular}{ll}
\hline \hline
Rashba-Dresselhaus anisotropy angle factor $\eta$ & $\pi/24$ 
\\
\hline
System size $(N_x,N_y)$ & $(512,1024)$ 
\\ 
\hline 
Lattice spacing $a$ & $(\pi/8\cos\!\eta)\times L_{\kappa}$
\\ 
\hline
Correlation amplitude $\gamma_0$ & $2\times E_{\kappa}^2 L_{\kappa}^2$
\\ 
\hline \hline
\end{tabular}
\end{tabular}
\caption{Parameters used for the simulations in the main text, where $L_{\kappa} = \kappa^{-1}$ and $E_{\kappa} = \hbar\kappa^2/(2m)$ are the natural length and energy scales introduced by the clean Hamiltonian. 
The spacings in momentum space are given by $\Delta k_{x,y} = 2\pi/(aN_{x,y})$, and
the disorder strength $W$ is determined from Eq.~(\ref{eq_SM:corr_amp_uniform_distrib}). }
\label{tab:simulation_params}
\end{table}
\renewcommand{\arraystretch}{1}

\end{document}